\documentclass[aip,jap,reprint]{revtex4-1} 
\newcommand{\APP}{$\mathrm{AP \rightarrow P}$ }
\newcommand{\PAP}{$\mathrm{P \rightarrow AP}$ }

\usepackage{amsmath}
\usepackage{amssymb}
\usepackage{latexsym}
\usepackage{indentfirst}
\usepackage{graphicx}
\usepackage{color}
\usepackage{bm}
\usepackage{multirow}

\begin{document}

\title{Temperature dependent nucleation and propagation of domain walls in a sub-100 nm perpendicularly magnetized Co/Ni multilayer}

\author{D.~B. Gopman}
\email{daniel.gopman@physics.nyu.edu}
\affiliation{Department of Physics, New York University, New
             York, NY 10003, USA}
\author{D. Bedau}
\affiliation{Department of Physics, New York University, New
             York, NY 10003, USA}
\affiliation{HGST San Jose Research Center,
             San Jose, CA 95135 USA}
\author{S. Mangin}
\affiliation{Institut Jean Lamour, UMR CNRS 7198 –Universit\'{e} de Lorraine, Nancy, France}
\author{E.~E. Fullerton}
\affiliation{CMRR, University of California at San Diego,
             La Jolla, CA 92093, USA}
\author{J.~A. Katine}
\affiliation{HGST San Jose Research Center,
             San Jose, CA 95135 USA}
\author{A.~D. Kent}
\affiliation{Department of Physics, New York University, New
             York, NY 10003, USA}

\begin{abstract}
We present a study of the temperature dependence of the switching fields in Co/Ni-based perpendicularly magnetized spin-valves. While magnetization reversal of all-perpendicular Co/Ni spin valves at ambient temperatures is typically marked by a single sharp step change in resistance, low temperature measurements can reveal a series of resistance steps, consistent with non-uniform magnetization configurations. We propose a model that consists of domain nucleation, propagation and annihilation to explain the temperature dependence of the switching fields. Interestingly, low temperature ($<30$ K) step changes in resistance that we associate with domain nucleation, have a bimodal switching field and resistance step distribution, attributable to two competing nucleation pathways.
\end{abstract}


\maketitle

Magnetic nanopillars with perpendicular magnetic anisotropy have garnered much attention for magnetic data storage applications.\cite{Mangin2006,Meng2006,Nakayama2008,Bedau2010,Ikeda2010,Worledge2011} The perpendicular anisotropy is well suited to device scaling, as the anisotropy is sensitive to surface and interface compositions rather than lateral geometry, which provides the shape anisotropy needed for in-plane magnetized devices. More important to spintronics applications, the critical current for switching a perpendicularly magnetized device is expected to be proportional to the magnetic anisotropy energy, which leads to lower currents required to reverse a nanoelement's magnetization for a given thermal stability.\cite{Mangin2009}

Nanopillar devices with a lateral size deep below 50~nm have recently been demonstrated.\cite{Gajek2012,Chun2012} Due to the large magnetostatic coupling between the perpendicularly magnetized free and polarizing layers, patterned devices smaller than 100~nm in diameter may not exhibit two remanent states in zero applied field. However, the polarizer may be replaced by a perpendicularly magnetized synthetic antiferromagnet composite to reduce this coupling and render the free layer bistable at zero field.\cite{Tudosa2010,Bandiera2010}

The thermal stability of a given perpendicularly magnetized nanoelement increases less rapidly and may even saturate above a critical lateral size, $L_D$, which depends on the exchange stiffnes and thickness of the individual nanoelement.\cite{Xiao2010,Sun2011,Chaves2013} Larger spin-valve devices exhibit complicated switching behavior. The free element in these devices typically can reverse through nucleation of a sub-volume whose size is comparable to $L_D$ followed by domain wall propagation.\cite{Bernstein2011} As large elements do not reverse coherently, the activation energy barrier is determined by the subvolume initiating the reversal. 

Measurements of the switching field and current in these devices at ambient temperatures may only be sensitive to the sub-volume nucleation event, as the thermal energy may overcome any domain wall pinning in the nanomagnet. This can explain why previous measurements of the thermally-assisted reversal behavior of nanomagnets appear to obey a simple thermal activation model over a single barrier, whose height differs from what is expected from a macrospin.\cite{Bedau2010b,Gopman2012} In this article, we study transitions between metastable intermediate device resistance states occurring during the reversal of a spin valve. Furthermore, we can conduct variable temperature measurements of the switching field between device resistance states in order to probe the activation barriers.



Our spin-valve nanopillars are magnetic multilayered films with strong uniaxial anisotropy perpendicular to the plane and have been described previously.\cite{Tudosa2010} The free layer is a Co/Ni multilayer and the polarizing layer is a synthetic antiferromagnet Co/Ni Co/Pd multilayer with a sufficiently higher coercive field to be considered fixed for all of our measurements. The layer  stack  is  composed  of  Ta(3)\slash Cu(35)\slash Pd(3)\slash [Co(0.3)\slash Pd(0.7)]$\times$4\slash Co(0.3)\slash Ru(0.9)\slash [Co(0.3)\slash Pd(0.7)]$\times$2 \slash [Co(0.15)\slash Ni(0.6)]$\times$2\slash Co(0.3)\slash Cu(4)\slash [Co(0.15)\slash Ni(0.6)]$\times$2\slash Co(0.3)\slash Pd(0.7)\slash Cu(15)\slash Ta(3)  (layer  thicknesses in  nanometers).  These  films  have  been  patterned  into 80~nm diameter circles by a process that combines electron beam and optical lithography. 

\begin{figure}[htb]
  \begin{center}
   \includegraphics[width=3.35 in,
    keepaspectratio=True]
   {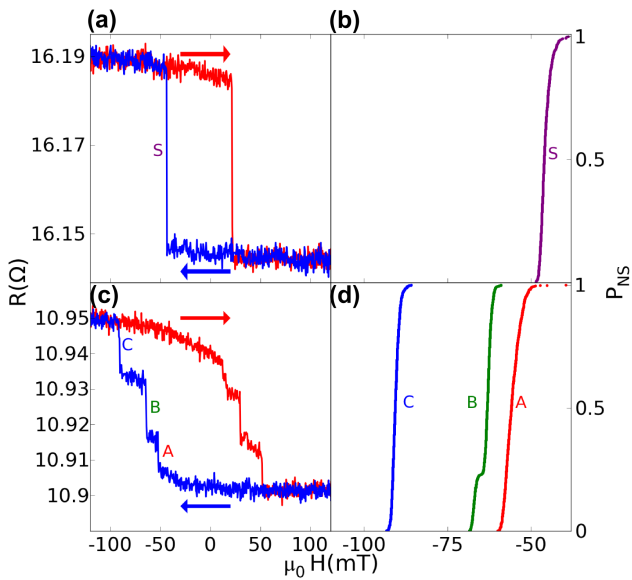}
  \end{center}
  \caption{\label{fig:F1} (a) Resistance vs perpendicular applied field hysteresis loop showing a single step (labeled $S$) switch. (b) Switching field distribution, probability of not switching, $\mathrm{P_{NS}}$, vs field for $S$ at room temperature obtained from 500 hysteresis loop measurements. (c) Hysteresis loop at T=12~K showing three resistance steps for the \PAP transition comprising an initial step $A$, followed by two intermediate transitions $B$ and $C$. Reversible changes in resistance precede the irreversible transitions \APP and \PAP. (d) Switching field distributions for the individual transitions at $12$~K  also taken from 500 hysteresis loops.}
\end{figure}

Quasistatic measurements of the sample magnetoresistance were taken within a cryostat using a lock-in detection scheme, with a 10~kHz excitation current of $I_{ac} = 50 \, \mu A $~rms (the room temperature, quasistatic switching current, $I_C = 0.5 \, \mathrm{mA} > I_{ac}$). Minor resistance hysteresis loops indicating the free layer reversal were recorded using a linear ramped magnetic field. Figure~\ref{fig:F1}(a) illustrates a typical resistance versus perpendicular applied field hysteresis loop exhibiting a single step at room temperature. By ramping the applied field several hundred times and recording the field at which the free layer reverses, defined by the corresponding step change in sample resistance, we sample from the switching field distribution, the probability of not switching, $P_{NS}$ for a given transition. The distribution for the single step (S) of the \PAP transition is displayed in Fig.~\ref{fig:F1}(b). We also present a hysteresis loop of this spin-valve at 12~K in Fig.~\ref{fig:F1}(c). We note that the transition \PAP now occurs in three steps, labeled on the figure as $A$, $B$ and $C$.

We associate the first step $A$ with a nucleation event. The second and third steps may then be associated with domain wall propagation and then its annihilation. However, we cannot distinguish between alternative interpretations, such as the intermediate switching events being multiple domain nucleation events. But multiple nucleation events would appear unlikely, as these 80~nm-diam spin valve devices appear too small to favor multi-domain configurations.  Figure~\ref{fig:F1}(d) shows the representative switching field distributions for the three steps. The inflection point in distribution $B$ near $P_{NS}=0.2$ is associated with different preceding nucleation events, which we discuss further below. We also note an increase in the resistance preceding the step $A$ for the \PAP transition branch, which we associate with a gradual fanning of the edge moments at fields below the nucleation field. We have tested that this change in the resistance is reversible with the applied field and the resistance change becomes more pronounced at lower temperatures. The particularly pronounced curling of the resistance in the \APP branch could be due to the larger fields at the free layer perimeter due to dipole fields from the polarizing layer when starting in the AP state.\cite{Gopman2012}

\begin{figure}[b]
  \begin{center}
   \includegraphics[width=3.35 in,
    keepaspectratio=True]
   {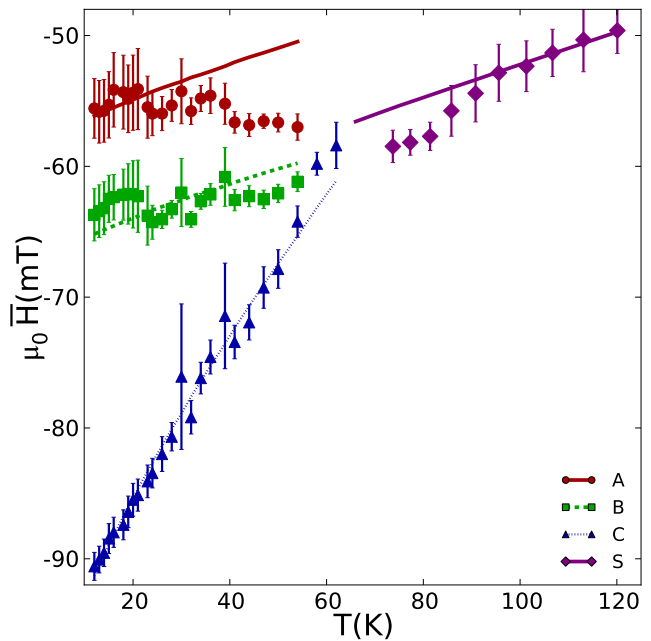}
  \end{center}
  \caption{\label{fig:F2} Evolution of the mean switching field $\mu_0\overline{H}$ with temperature $T$ for transitions of the \PAP switching branch comprising three step changes in resistance at low temperature: $A$ (red circles), $B$ (green squares) and $C$ (blue triangles). These three trendlines intersect around T=80~K, above which only a single-step resistance change, $S$ (purple diamonds), is evident. Error bars reflect the variance of the transitions at each temperature. Solid lines reflect thermal activation prediction for $\mu_0\overline{H}(T)$ with best-fit parameters listed in Table~\ref{Tab:FitParams}.}
\end{figure}

The mean switching field for steps $A$, $B$ and $C$ were recorded for temperatures from 12~K to 120~K from 500 minor hysteresis curves taken at each temperature. We show the evolution of the mean transition field for each of these processes comprising a \PAP switch in Fig.~\ref{fig:F2}. Error bars reflect the variance of the switching distribution at a given temperature. For temperatures below 80~K, we note the gradual trend of $A$ and $B$ steps (red circles and green squares, respectively) compared to the more rapid decrease in the switching field for step $C$ (blue triangles). This difference in slopes reflects the lower barrier for transition $C$ relative to the other transitions, as the slope varies in proportion to the ratio of thermal energy to the activation barrier. Around 80~K, the mean switching trendline for the three transitions intersects, above which hysteresis minor loops only reflect a single step change in the resistance, $S$ (purple diamonds), as in Fig.~\ref{fig:F1}(a).

We analyze our temperature-dependent switching data assuming thermal activation over a field-dependent energy barrier within the N\'eel-Brown model.\cite{Neel1949,Brown1963} At fixed temperatures and fields one can define an Arrhenius rate of escape as $\Gamma(H) = \Gamma_0 e^{- E ( H ) /k_BT}$, where $\Gamma_0$ is the attempt frequency and $k_B$ is Boltzmann's constant. The form of the energy barrier, $E(H) = E_0(1-H/H_{c0})^\eta$, where $\eta$=1.5, $E_0$ is the energy barrier at zero field, and $H_{c0}$ is the zero-temperature coercive field.\cite{Victora1989,Coffey1995} The cumulative probability to remain in a metastable magnetization state under finite field, $\mu_0H$, is $\exp{[-\frac{1}{v}\int_0^{H}\,\Gamma(H')dH']}$, where $v$ is the ramp rate of the magnetic field (100~mT/s).\cite{Kurkijarvi1972,Garg1995}. An approximate expression for the mean transition field is given by:\begin{eqnarray}
&& \overline{H} \cong H_{c0} \left( 1 - \left[ \frac{1}{\xi} \log \left( \frac{ \Gamma _0 H_{c0} }{ \eta v \xi \varepsilon ^{ \eta - 1} } \right)  \right]^{1 / \eta} \right),
\label{Energy}
\end{eqnarray}
where $\xi = E_0 / k_BT$ is referenced to $T=300~K$ and $\varepsilon = (1 - H/H_{c0})$. We obtain a reasonably fit for the mean transition field (lines of best-fit in Fig.~\ref{fig:F2}) for the individual transitions up to approximately 80~K, after which the reversal process collapses onto a single step, which we fit for higher temperatures.

\begin{table}[h]
	\noindent{}\begin{tabular}{c|c||c||c||c|l}	
	\cline{2-5}
	& $A$ & $B$ & $C$ & $S$\\ \cline{1-5}
	\multicolumn{1}{|c|}{$\xi$} & 50 (20) & 16.5 & 9 & 48 &\\ \cline{1-5}
	\multicolumn{1}{|c|}{$| \mu_0 H_{c0}|$(mT)} & 60 (62) & 69 & 97 & 69  &   \\ \cline{1-5}
	\end{tabular}
\caption{Barrier heights $\xi$ and zero-temperature coercive fields $\mu_0 H_{c0}$ for \PAP transitions: $A$, $B$, $C$ and $S$ (no intermediate steps). Parameters in parentheses in column $A$ reflects the best-fit parameters of a second switching mode ($A'$) in the distribution fields for this transition. At temperatures above approximately 80~K, all three mean switching fields collapse onto the trendline for transition $S$.}
\label{Tab:FitParams}
\end{table}

From the best-fit lines of the transition fields in Fig.~\ref{fig:F2} we extract $E_0$, the barrier height, and $H_{c0}$, the transition field at zero temperature. These parameters are summarized in Table~\ref{Tab:FitParams}. As could be inferred by the steeper slope for step $C$, the barrier height for the annihilation process (9~$k_BT$) is smaller (by a factor of 4) than the nucleation ($A$) and propagation ($B$) barriers. The nucleation field cannot be adequately fit by the model since it reflects more than one competing process, which we will address below. Calculating the lifetime of a metastable state as the inverse of the escape rate ($\tau = 1 / \Gamma$), we easily see that the ridge preceding domain wall annihilation disappears above 80~K as the lifetime dips below 1~ms, which is the sampling frequency of our quasistatic measurements. We have a similar result for the pinning that occurs prior to the propagation transition, whose energy barrier (16.5~$k_BT$) and critical field (69~mT) are easily overcome by thermal activation above 80~K. Finally, the barrier corresponding to the single step that persists to high temperatures (48~$k_BT$) is smaller than the macrospin barrier height, $E_B = \mu_0 M_S H_K V / 2 \approx 140 \, k_BT$ ($M_S = 600 \, \mathrm{kA/m}$ and $H_K = 0.2$~T), which is consistent with a sub-volume activation, accompanied by domain propagation. The magnitude of sub-volume activation barrier can be estimated using the model developed by Sun \textit{et al.}\cite{Sun2011}, $E_B \sim 4 \pi A_{ex} t$, which considers the exchange constant $A_{ex}$ and thickness $t$ of our free layer. Taking a weighted sum of the exchange constants for Co\cite{Tannenwald1961} and Ni,\cite{Nose1961} $A_{ex} = 0.93 \times 10^{-11}$~J/m and $t = 1.8$~nm, we estimated a sub-volume barrier of 67~$k_BT$, which is reasonably close to the barrier for our single step process.

\begin{figure}[b]
  \begin{center}
   \includegraphics[width=3.35 in,
    keepaspectratio=True]
   {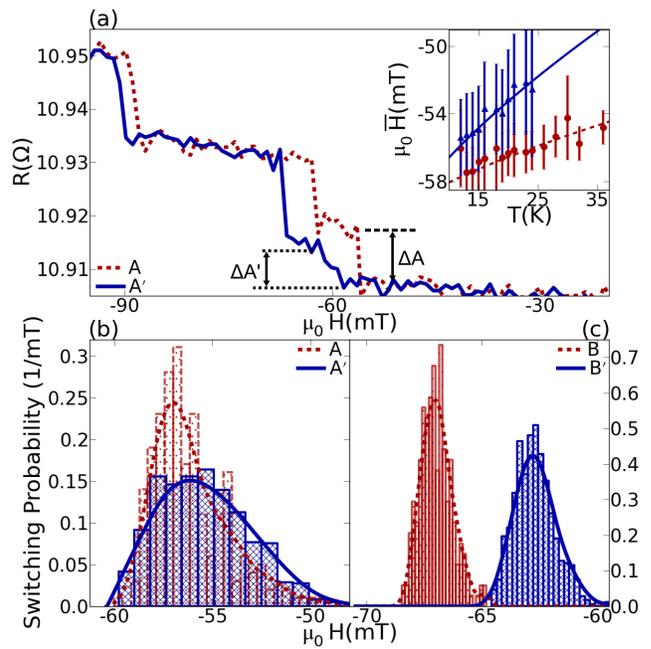}
  \end{center}
  \caption{\label{fig:F3} Competing initial resistance steps at T=12~K. (a) Typical R vs H hysteresis curves ($A$, $A'$) showing distinct initial resistance jumps corresponding to the steps $A$ (broken red line) and $A'$ (solid blue line) (Inset: evolution of the mean transition field for $A$ and $A'$ under varied temperature). (b) Switching Probability versus Field for resistance steps $A$  and $A'$. (c) Switching Probability versus Field for propagation processes $B$  and $B'$. Association with $A$  and $A'$ correlates with initial resistance change $\mathrm{\Delta R}$ and transition field for $B$  and $B'$ falling within first or second distribution mode.}
\end{figure}

We will now consider the values in parentheses in the nucleation column of Table~\ref{Tab:FitParams}. In the inset of Fig.~\ref{fig:F1}(b), distribution $B$ exhibits a plateau followed by a second dip in the switching field distribution. This signature of a bimodal distribution in the propagation switching transition appears at temperatures below 25~K and may be a consequence of two competing nucleation processes, each with its own resistance signature. We demonstrate this phenomenon in Fig.~\ref{fig:F3}(a) with two \PAP transition curves $A$ and $A'$, whose unique resistance changes $\Delta A$ and $\Delta A'$ distinguish the two different switching processes. The switching histograms in Fig.~\ref{fig:F3}(b) illustrate the distinct distributions underlying processes $A$ and $A'$, and Fig.~\ref{fig:F3}(c) depicts the temperature dependence of the mean nucleation field for the two processes. By separating the two modes for $A$ and $A'$ (accordingly with $B$ and $B'$) according to their resistance signatures, we present the best-fit parameters $E_0$ and $H_{c0}$ of the main mode that exists up to 80~K as well as the competing (primed) mode in parenthesis in Table~\ref{Tab:FitParams}. 

The competing reversal processes underlying the two initial steps in resistance at low temperatures requires careful attention. There is a significant difference ($\mathrm{100 \%}$) between the barriers of the two competing modes $A$ and $A'$. If the two barriers represented competing pathways from an identical initial state, we should expect the mode with the lowest barrier ($A'$) to dominate the switching behavior. Instead, we see the large barrier process dominate the switching at elevated temperatures (T $>$ 25~K) as we saw in the inset of Fig.~\ref{fig:F3}(a). We also note that the reversible region of the R vs H hysteresis loop 
that precedes the switching transition is extended over a broader field range at lower temperatures (Fig.~\ref{fig:F1}(a,c)). It may be that the free layer transitions into a canted or fanned state, which could lead to significantly different switching behavior. We also cannot eliminate the possibility that the free layer is not completely reversed following each \APP transition, which may lower the barrier or change the process defined by $A'$. There is no cross-over between distributions after the nucleation step, which could indicate different domain wall types, prohibiting transitions between the two states $B$ and $B'$. We conclude that the last non-uniform state $C$ is identical for both nucleation modes based upon the single-mode distribution for this step.

\begin{figure}[htb]
  \begin{center}
   \includegraphics[width=3.35 in,
    keepaspectratio=True]
   {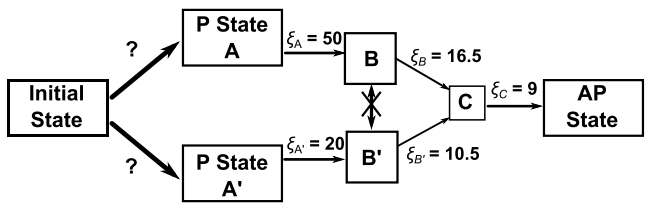}
  \end{center}
  \caption{\label{fig:logic} Diagram of the possible reversal pathways of our spin-valve device at low temperature, illustrating two competing nucleation events. An initial state leading to two indistinguishable $P$ states $A$,$A'$ evolve with distinct barrier heights into states $B$,$B'$. These states both evolve into $C$, which in turn transitions into the final ($AP$) state.}
\end{figure}

We summarize the series of steps leading to a \PAP transition at low temperature in Fig.~\ref{fig:logic}. Starting from an initial state following an \APP transition, the free layer may exist in either state $A$ or $A'$, depending on a pre-selection event that we cannot resolve with resistance measurements. The subsequent activation barrier for the $A \rightarrow B$ or $A' \rightarrow B'$ transitions are known and result in distinct transition field distributions. The states $B$ and $B'$ are on different pathways due to their distinct histories and present two non-overlapping switching distributions. Both $B$ and $B'$ lead to state $C$, whose transition to the final $AP$ state is identical for both initial competing pathways.

We have demonstrated the temperature dependence of nucleation and propagation fields in a Co/Ni nanomagnet. The switching distributions for the individual transitions are well described by thermal activation over a single energy barrier describing the subvolume activation energy or depinning energy of that transition. While the intermediate states disappear from our quasistatic measurements at elevated temperatures, this may reflect the thermal activation process that proceeds during each hysteresis cycle, but on a much shorter time scale than our approximately transport measurements can resolve ($\sim 1$ ms). The intermediate resistance states of a Co/Ni nanomagnets at low temperatures reveal details of a multistep reversal.

We also presented evidence for competing reversal processes at low temperatures. That this competition sets in at the lowest temperatures could suggest that as the thermal energy becomes comparable to the difference between two or more nucleation modes, the magnet can choose from the different modes during each hysteresis cycle. The subsequent behavior or the magnet (e.g. pinning, propagation) is then apparently frozen in by this choice. This complicated behavior is evidence of an energy landscape for switching, which should be relevant for understanding the switching of spintronics devices even at temperatures and timescales that no longer allow for the detection of intermediate non-uniform magnetization states.


\section*{Acknowledgments}
This research was supported at NYU by NSF Grant Nos. DMR-1006575 and NSF-DMR-1309202, as well as the Partner University Fund (PUF) of the Embassy of France. Research at UL supported by ANR-10-BLANC-1005 ``Friends'', the European Project (OP2M FP7-IOF-2011-298060) and the Region Lorraine. Work at UCSD supported by NSF Grant No. DMR-1008654. 

\bibliographystyle{apsrev}



\end{document}